\documentclass[12pt,fleqn]{article}
\usepackage{amsmath,amssymb,amsthm,tikz,indentfirst,graphicx,mathrsfs,enumerate,epstopdf,caption,subcaption,float,authblk}
\usepackage[square,comma,numbers,sort&compress]{natbib}
\usepackage[colorlinks]{hyperref}
 \theoremstyle{plain}
\newtheorem{theorem}{Theorem}[section]
\newtheorem{corollary}[theorem]{Corollary}
\newtheorem{lemma}[theorem]{Lemma}
\newtheorem{assume}[theorem]{Assumption}
\newtheorem{prop}[theorem]{Proposition}
 \theoremstyle{remark}

 \theoremstyle{definition}
 
\newtheorem{problem}[theorem]{Riemann--Hilbert Problem}
\usepackage[left=2.5cm,right=2.5cm,bottom=2.8cm,top=2.8cm]{geometry}

\newcommand{\res}{\operatorname{Res}\ }
\newcommand{\ii}{\mathrm{i}}
\newcommand{\dd}{\mathrm{d}}
\newcommand{\e}{\mathrm{e}}
\numberwithin{equation}{section}
\allowdisplaybreaks[2]
\begin{document}
\title{ Riemann--Hilbert method to the  Ablowitz--Ladik equation: higher-order  cases}

\author[1]{Huan Liu}
\author[2]{Jing Shen}
\author[1]{Xianguo Geng \thanks{Correspondence. E-mail: xggeng@zzu.edu.cn}}
\affil[1]{School of Mathematics and Statistics, Zhengzhou University, Zhengzhou, Henan 450001, People's  Republic  of China}
\affil[2]{School  of  Sciences,  Henan  University  of  Technology,  Zhengzhou, Henan  450001,  People's  Republic  of China}

\renewcommand*{\Affilfont}{\small\it}
\renewcommand{\Authands}{, }
\date{}

  \maketitle
   \begin{abstract}
   We focus on the Ablowitz-Ladik equation on the zero background, specifically considering the scenario of $N$ pairs of multiple poles. Our first goal was to establish a mapping between the initial data and the scattering data, which allowed us to introduce a direct problem by analyzing the discrete spectrum associated with $N$ pairs of higher-order zeros. Next, we constructed another mapping from the scattering data to a $2\times2$ matrix Riemann--Hilbert problem  equipped with several residue conditions set at $N$ pairs of multiple poles. By characterizing the inverse problem on the basis of this Riemann--Hilbert problem, we are able to derive higher-order soliton solutions in the reflectionless case.
  \end{abstract}

  \textbf{Keywords:} Ablowitz--Ladik equation;   Riemann--Hilbert problem; higher-order  soliton
\newpage

\section{Introduction}
The Ablowitz--Ladik equation \cite{Ablowitz1975,ablowitz2004discrete,Ablowitz1976Nonlinear}
\begin{equation}\label{eq:dfnls}
  \ii\frac{\dd q_n}{\dd t}=q_{n+1}-2q_n+q_{n-1}+|q_n|^2(q_{n+1}+q_{n-1}),
\end{equation}
 is known as a discrete version of the focusing nonlinear Schr\"odinger equation.
 It has attracted considerable attention in various physical systems, including  Heisenberg spin chains \cite{Ishimori1982,Papanicolaou1987}, self-trapping on a dimer \cite{Kenkre1986},  the dynamics of anharmonic lattices \cite{Takeno1990}.
The Ablowitz-Ladik equation has been studied extensively from different aspects, such as the initial-boundary problem \cite{XIA201827}, inverse scattering transform on the nonzero background \cite{Ablowitz_2007}, integrable decomposition and quasi-periodic solutions\cite{GengDZ,GengDC,ChenP}, long-time asymptotics \cite{Yamane2015} and $l^2$-Sobolev space bijectivity \cite{CHEN2023133565}.

In the framework of inverse scattering transform, soliton solutions of integrable equations are related to the poles of the transmission coefficient. For example, a one-soliton solution arises from a pair of conjugate simple poles, while an
$N$-soliton solution corresponds to $N$ pairs of conjugate simple poles. Therefore,  it is natural to investigate the behavior when these poles become higher-order instead of simple. Higher-order poles solutions of several integrable equations have been studied, including the nonlinear Schr\"odinger  equation \cite{ZS1972,Olmedilla1987,Aktosun2007,ZhangHe2020,Schiebold2017,BilmanBW}, modified Korteweg--de Vries equation \cite{Wadati1982,ZhangTao2020},  sine-Gordon equation \cite{Wadati1984},   $N$-wave system \cite{SYang2003,ShYang2003}, derivative nonlinear Schr\"odinger equation \cite{ZY2020},  modified short-pulse system \cite{LQL2022, LL2023}.

 However, there is little attention on discrete integrable systems with arbitrary order poles. Unlike continuous integrable systems, the spectral problem of discrete integrable equations always involves symmetry between inside and outside the circle. This symmetry is the greatest challenge that hinders the study of higher-order pole solutions of discrete integrable equations. In recent references \cite{CF2022,CFH2023}, the authors considered the discrete sine-Gordon equation and the discrete mKdV equation with simple and double poles. However, these approaches do not seem to work well in investigating higher-order poles uniformly. To the best of our knowledge, the inverse scattering transformation for the Ablowitz--Ladik equation \eqref{eq:dfnls} with arbitary order poles has not been reported. Therefore, this paper  aims to construct a matrix Riemann--Hilbert (RH) problem with several residue conditions at $N$ pairs of multiple poles, and provide an inverse scattering transformation for the study of higher-order soliton solutions of the Ablowitz--Ladik equation.

The remainder of this paper is organized as follows: In Section \ref{sec:dir}, we establish  a mapping between the initial data and the scattering data and analyse the discrete spectrum associated with  $N$ pairs of higher-order zeros. In Section \ref{inverse}, we construct another mapping from the scattering data to a $2\times2$ matrix RH problem  equipped with several residue conditions set at $N$ pairs of multiple poles.  In Section \ref{less}, we  derive  the higher-order pole soliton solutions in the reflectionless case.

\section{ Direct scattering problem }\label{sec:dir}

The Ablowitz--Ladik  equation \eqref{eq:dfnls}  can be interpreted as a compatibility condition for the system of simultaneous linear equations in a Lax pair \cite{Ablowitz1975,ablowitz2004discrete,Ablowitz1976Nonlinear}:
\begin{equation}\label{Lax}
  \phi_{n+1}=\mathbf{U}_n\phi_n,\quad \frac{\dd\phi_n}{\dd t}=\mathbf{V}_n\phi_n,
\end{equation}
with
\begin{equation}
\begin{aligned}
  &\mathbf{U}_n:=\mathbf{U}_n(\lambda;t)=\Lambda+\mathbf{Q}_n,\\ &\mathbf{V}_n:=\mathbf{V}_n(\lambda;t)=-\frac{\ii}{2}(\lambda-\lambda^{-1})^2\sigma_3+\ii\sigma_3\left[\mathbf{Q}_n\mathbf{Q}_{n-1}-\mathbf{Q}_n\Lambda^{-1}+\mathbf{Q}_{n-1}\Lambda\right],\label{uvdef}\\
&\Lambda=\begin{pmatrix}
  \lambda &\mathbf{0}\\
  \mathbf{0}&\lambda^{-1}
\end{pmatrix}, \quad \mathbf{Q}_n=\begin{pmatrix}
  0&q_n\\
   -\bar q_n&0
\end{pmatrix},\quad\sigma_3=\begin{pmatrix}
                              1 & 0 \\
                              0 & -1
                            \end{pmatrix},
\end{aligned}
\end{equation}
where $\phi_n:=\phi_n(\lambda;t)$ is a $2\times 2$ matrix-valued function of $n$, $t$ and the spectral parameter $\lambda\in\mathbb C$, `` $\bar{}$ " denotes the complex conjugate. In other words, the   Ablowitz--Ladik equation \eqref{eq:dfnls}  is equivalent to the zero-curvature equation
$\frac{\dd \mathbf{U}_n}{\dd t}+\mathbf{U}_n\mathbf{V}_n-\mathbf{V}_{n+1}\mathbf{U}_n=\mathbf{0}$.

\subsection{Jost solutions and scattering matrix}
In this paper, to simplify the analysis, the time variable $t$ will be omitted since the time evolution of the scattering data is explicitly considered. Suppose that $q_n$ decays rapidly for large $n$.
We seek two Jost solution matrices  $\varphi_n(\lambda)$ and $\psi_n(\lambda)$ solving Eq.\,\eqref{Lax} for $n\in \mathbb{Z}$ that satisfies the boundary conditions on $|\lambda|=1$
\begin{equation}\label{eq:psiasy}
\begin{aligned}
   & \varphi_n(\lambda)= \Lambda^n+o(1), \quad n\rightarrow -\infty,\\
    &  \psi_n(\lambda)= \Lambda^n+o(1), \quad n\rightarrow +\infty.
\end{aligned}
\end{equation}
Making the normalization
\begin{equation}\label{eq:mJost}
  \Phi_n(\lambda)=\varphi_n(\lambda)\Lambda^{-n},\quad \Psi_n(\lambda)=\psi_n(\lambda)\Lambda^{-n},
\end{equation}
we obtain
\begin{subequations}\label{eq:mulax}
\begin{align}
 & \Phi_{n+1}(\lambda)=\Lambda\Phi_n(\lambda)\Lambda^{-1}+\mathbf{Q}_n\Phi_n(\lambda)\Lambda^{-1},\label{eq:mumlax}\\
  &(1+|q_n|^2)\Psi_n(\lambda)=\Lambda^{-1}\Psi_{n+1}(\lambda)\Lambda-\mathbf{Q}_n\Psi_{n+1}(\lambda)\Lambda.\label{eq:muplax}
\end{align}
\end{subequations}
Denote \begin{equation}
  \delta^+_n=\prod_{k=n}^{+\infty}(1+|q_k|^2),\quad \delta^-_n=\prod_{k=-\infty}^{n-1}(1+|q_k|^2),
\end{equation} and assume that $\delta=\delta^+_n\delta^-_n<\infty$. Indeed, it is easy to check that $\delta$ is a conserved quantity which does not depend on $t$. Let $ \tilde\Psi_n(\lambda)=\delta^+_n\Psi_n(\lambda)$,
Eq.\,\eqref{eq:muplax} can be rewritten as
\begin{equation}
  \tilde\Psi_n(\lambda)=\Lambda^{-1}\tilde\Psi_{n+1}(\lambda)\Lambda-\mathbf{Q}_n\tilde\Psi_{n+1}(\lambda)\Lambda.
\end{equation}
Noting the asymptotics
\begin{equation}\label{eq:muasy}
\begin{aligned}
   & \Phi_n(\lambda)= \mathbf{I}+o(1), \quad n\rightarrow -\infty,\\
    &  \tilde\Psi_n(\lambda)= \mathbf{I}+o(1), \quad n\rightarrow +\infty,
\end{aligned}
\end{equation}
we know that $ \Phi_n(\lambda)$  and $\tilde\Psi_n(\lambda)$ satisfy the following summation equation,
\begin{subequations}\label{eq:mupmin}
\begin{align}
  \Phi_n(\lambda)=&
   \mathbf{I}+\sum_{k=-\infty}^{n-1}\Lambda^{n-1-k}\mathbf{Q}_k\Phi_k(\lambda)\Lambda^{k-n},\label{eq:mupmin1}\\
\tilde\Psi_n(\lambda)=&
   \mathbf{I}-\sum_{k=n}^{+\infty}\Lambda^{n-k}\mathbf{Q}_k\tilde\Psi_{k+1}(\lambda)\Lambda^{k-n+1}.\label{eq:mupmin2}
\end{align}
\end{subequations}

Let $D_\pm=\left\{\lambda\in\mathbb{C}\big|\ \vert \lambda\vert\gtrless 1\right\}$, $\Gamma=\{\lambda\in\mathbb{C}\big|\ \vert \lambda\vert=1\}$.
   For a $2\times 2$ matrix $\mathbf{A}_n$, $\mathbf{A}_{n,j}$ represents the $j$-th column, $\mathbf{A}_{n,ij}$ represents the $(i,j)$-element.

\begin{theorem}\label{Th:ana}
  Suppose that $ \sum_{n=-\infty}^{+\infty}|q_n|<\infty$, $\Phi_n(\lambda)$ and $\Psi_n(\lambda)$ are well-defined in  $\Gamma$,  the first column of  $\Phi_n(\lambda)$ and the second column  of  $\Psi_n(\lambda)$, denoted as  $\Phi_{n,1}(\lambda)$ and $\Psi_{n,2}(\lambda)$ respectively,  can be analytically continued onto $D_+$, while the second column  of  $\Phi_n(\lambda)$ and the first column of  $\Psi_n(\lambda)$, denoted as $\Phi_{n,2}(\lambda)$ and $\Psi_{n,1}(\lambda)$  respectively,  can be analytically continued onto $D_-$.
  Moreover, the functions $\varphi_n(\lambda)$ and $\psi_n(\lambda)$ have the same analyticity  properties as $\Phi_n(\lambda)$ and $\Psi_n(\lambda)$, respectively.
\end{theorem}

Since $\det(\mathbf{U}_n)=1+|q_n|^2$, then for $\lambda\in\Gamma$,
\begin{equation}
\begin{aligned}
    &\det[\Phi_n(\lambda)]=(1+|q_{n-1}|^2)\det[\Phi_{n-1}(\lambda)]=\delta^-_n\lim_{k\rightarrow -\infty}\det\Phi_k(\lambda)=\delta^-_n, \\
    & \det[\Psi_n(\lambda)]=(1+|q_n|^2)^{-1}\det[\Psi_{n+1}(\lambda)]=(\delta^+_n)^{-1}\lim_{k\rightarrow +\infty}\det\Psi_k(\lambda)=(\delta^+_n)^{-1}.
\end{aligned}
\end{equation}
Additionally,
 \begin{equation}\label{eq:psidet}
    \det[\varphi_n(\lambda)]=\delta^-_n,\quad   \det[\psi_n(\lambda)]=(\delta^+_n)^{-1},\quad \lambda\in\Gamma.
 \end{equation}
In other words, both $\psi_n(\lambda)$ and $\varphi_n(\lambda)$ are the nonsingular solution matrices. Besides, they satisfy  the same differential equation (the spatial part of Lax pair \eqref{Lax}), we can infer that there exists a $2\times2$ matrix $\mathbf{s}(\lambda)$  independent of $n$ such that
\begin{equation}\label{eq:Jostjump}
  \varphi_n(\lambda)=\psi_n(\lambda)\mathbf{s}(\lambda),\quad \lambda\in\Gamma.
\end{equation}
Moreover, evaluating Eq.\,\eqref{eq:Jostjump}  as $ n\rightarrow +\infty$ and considering Eqs.\,\eqref{eq:mJost} and \eqref{eq:mupmin1}, we can obtain that the scattering matrix $s(\lambda)$ can be represented as:
\begin{equation}\label{eq:sin}
   \mathbf{s}(\lambda)=\mathbf{I}+\sum_{k=-\infty}^{+\infty}\Lambda^{-1-k}\mathbf{Q}_k\Phi_k(\lambda)\Lambda^{k}.
\end{equation}
We observe that the scattering matrix $s(\lambda)$ is determined by  the data $\{q_n\}_{n=-\infty}^{+\infty}$.
\subsection{Symmetries and asymptotics}

The Jost eigenfunctions and the scattering matrix satisfy certain symmetry relations that arise from the symmetries of the Lax pair. These symmetries play a crucial role in solving the inverse problem and finding explicit solutions of the   Ablowitz--Ladik equation \eqref{eq:dfnls}.  The Lax pair possesses two symmetrie: $\lambda\mapsto\bar{\lambda}^{-1}$ and $\lambda\mapsto-\lambda$. Direct calculations lead to this the following proposition.
\begin{prop}
  If $\phi_n(\lambda)$ is a fundamental matrix  solution of the Lax pair \eqref{Lax}, so are  $\sigma_2\overline{\phi_n(\bar\lambda^{-1})}$ and $\sigma_3\phi_n(-\lambda)$, where
  $
    \sigma_2=\begin{pmatrix}
      0&-\ii\\\ii&0
    \end{pmatrix}$, $\sigma_3=\begin{pmatrix}
      1&0\\0&-1
    \end{pmatrix}$.
\end{prop}
By substituting either one of the fundamental matrix solutions  into the above symmetry relation and combining it with the asymptotics \eqref{eq:psiasy}, we can infer that
\begin{subequations}\label{eq:Jostsym}
  \begin{alignat}{1}
    &\varphi_n(\lambda)=\sigma_2\overline{\varphi_n(\bar\lambda^{-1})}\sigma_2,\quad \psi_n(\lambda)=\sigma_2\overline{\psi_n(\bar\lambda^{-1})}\sigma_2,\label{eq:Jostsym1}\\
&
  \varphi_n(\lambda)=(-1)^n\sigma_3\varphi_n(-\lambda)\sigma_3,\quad \psi_n(\lambda)=(-1)^n\sigma_3\psi_n(-\lambda)\sigma_3.\label{eq:Jostsym2}
  \end{alignat}
\end{subequations}
  The above symmetry \eqref{eq:Jostsym} holds only for $\lambda\in\Gamma$ because the columns of $\varphi_n(\lambda)$ and $\psi_n(\lambda)$ are not all analytic in the same domain. We can expand the symmetry \eqref{eq:Jostsym} column-wise to  the corresponding regions of analyticity.

  Moreover,  the symmetry also affects the scattering matrix. By combining the symmetry \eqref{eq:Jostsym} of the eigenfunctions  with the scattering relation \eqref{eq:Jostjump}, we can derive two symmetries  of the scattering matrix:
   \begin{equation}\label{eq:ssym}
    \mathbf{s}(\lambda)=\sigma_2\overline{\mathbf{s}(\bar\lambda^{-1})}\sigma_2,\quad  \mathbf{s}(\lambda)=\sigma_3\mathbf{s}(-\lambda)\sigma_3.
  \end{equation}
 Therefore, the scattering matrix $\mathbf{s}(\lambda)$ can be written in the form
  \begin{equation}
    \mathbf{s}(\lambda)=\begin{pmatrix}
      a(\lambda)&-\overline{b(\bar\lambda^{-1})}\\
      b(\lambda)&\overline{a(\bar\lambda^{-1})}
    \end{pmatrix},
  \end{equation}
  where
  \begin{equation}\label{sym:ab}
  a(\lambda)=a(-\lambda),\quad b(\lambda)=-b(-\lambda).
\end{equation}
Based on Theorem \ref{Th:ana}  and the  integral representation  given in Eq.\,\eqref{eq:sin}, we find that $a(\lambda)$ and $b(\lambda)$ are well-defined within the region  $\Gamma$. By taking the determinants of both sides of Eq.\,\eqref{eq:Jostjump} and recalling Eq.\,\eqref{eq:psidet}, we can infer that
\begin{equation}
  |a(\lambda)|^2+|b(\lambda)|^2=\delta,\quad \lambda\in\Gamma.
\end{equation}
  The scattering coefficients $a(\lambda)$ and  $b(\lambda)$ can be expressed as determinants of columns of $\varphi_n(\lambda)$ and  $\psi_n(\lambda)$:
 \begin{subequations}\label{eq:sS}
 \begin{alignat}{2}
 &a(\lambda) =\delta^+_n\operatorname{det}[\varphi_{n,1}(\lambda),\psi_{n,2}(\lambda)],\qquad&& \lambda\in  \Gamma,\label{eq:sSa}\\
 &\overline{a(\bar\lambda^{-1})}=\delta^+_n\operatorname{det}[\psi_{n,1}(\lambda),\varphi_{n,2}(\lambda)],\qquad &&\lambda\in   \Gamma,\label{eq:sSb}\\
 &b(\lambda) =\delta^+_n\operatorname{det}[\psi_{n,1}(\lambda),\varphi_{n,1}(\lambda)],\qquad&& \lambda\in  \Gamma,\\
& \overline{b(\bar\lambda^{-1})}=\delta^+_n\operatorname{det}[\psi_{n,2}(\lambda),\varphi_{n,2}(\lambda)],\qquad &&\lambda\in \Gamma.
  \end{alignat}
  \end{subequations}
Furthermore, it follows from Eqs.\,\eqref{eq:sSa} and \eqref{eq:sSb} that $a(\lambda)$ and $\overline{a(\bar\lambda^{-1})}$ can be analytically continued  to the region $D_+$ and $D_-$, respectivley.

In the inverse problem, the following reflection coefficient will appear,
\begin{equation}\label{eq:rho}
 \gamma(\lambda)=\frac{b(\lambda)}{a(\lambda)},\quad \lambda\in\Gamma.
  \end{equation}
  Also, it follows from Eq.\,\eqref{sym:ab} that
  \begin{equation}\label{eq:rhoh}
 \gamma(\lambda)=-\gamma(-\lambda).
  \end{equation}

In the context of the IST, it is necessary to understand the asymptotic behavior of the Jost eigenfunctions and the scattering matrix as the spectral parameter tends to infinity and zero.
It is straightforward to substitute the Wentzel--Kramers--Brillouin  expansions of the columns of the modified Jost solutions into Eq.\,\eqref{eq:mulax} and explicitly compute the first few terms of these expansions   as $\lambda$ approaches infinity and zero in the appropriate regions.
\begin{prop}\label{prop:casyeg}
  As $\lambda\in D_+$ and $\lambda\rightarrow\infty$,
  \begin{equation}\label{eq:asyeg}
  (\varphi_{n,1}(\lambda),\psi_{n,2}(\lambda))=\begin{pmatrix}
    1& -\lambda^{-1}(\delta^+_n)^{-1}q_n \\
    -\lambda^{-1}\bar q_{n-1} &(\delta^+_n)^{-1}
  \end{pmatrix}+O(\lambda^{-2}),
\end{equation}
Similarly, as $\lambda\in D_-$ and $\lambda\rightarrow 0$,
 \begin{equation}\label{asyeg0}
   (\psi_{n,1}(\lambda),\varphi_{n,2}(\lambda))=\begin{pmatrix}
    (\delta^+_n)^{-1}&\lambda q_{n-1} \\
   \lambda(\delta^+_n)^{-1}\bar q_n &1
  \end{pmatrix}+O(\lambda^2).
\end{equation}
\end{prop}
In addition, combining the determinant form \eqref{eq:sSa} of the scattering coefficient $a(\lambda)$ with the asymptotics in \eqref{eq:asyeg} yields \begin{equation}
a(\lambda)=1+O(\lambda^{-2}),\quad\lambda \rightarrow \infty.
\end{equation}

\subsection{Discrete spectrum}
\begin{prop}\label{prop:psi}
If $ w$ is a zero of $a(\lambda)$ with multiplicity $m+1$, then there exist $m+1$ complex-valued constants $b_0,b_1,\ldots, b_m (\,b_0\neq 0)$ such that
 \begin{equation}\label{eq:psid}
  \frac{ \varphi_{n,1}^{(h)}( w)}{n!}=\sum_{j+k=h\atop j,k\geqslant 0}\frac{ b_j\psi_{n,2}^{(k)}(w)}{j!k!},
 \end{equation}
 holds for all $n\in\mathbb{Z}$ and each $h\in\{0,\ldots,m\}$, where   $f^{(s)}(w)$ denotes $\frac{\dd^sf(\lambda)}{\dd \lambda^s}|_{\lambda=w}$ for a function $f(\lambda)$.
\end{prop}
\begin{proof}
   When $h=0$, it can be deduced from Eq.\,\eqref{eq:sSa} that the vectors $\varphi_{n,1}(w)$ and $\psi_{n,2}(w)$ are linearly dependent. However, both vectors are nonzero according to Eq.\,\eqref{eq:psiasy}. Therefore, there must exist a nonzero complex-valued constant $b_0$ such that $\varphi_{n,1}(w)=b_0\psi_{n,2}(w)$.

   Let us assume that the proposition holds for all $h<j$, where $j$ is  a fixed positive integer not exceeding $m$. This means that there exist complex-valued constants $b_0,\ldots, b_{j-1}$ such  that
    for each $h\in\{0,\ldots,j-1\}$,
 \begin{equation}\label{ls1}
  \frac{ \varphi_{n,1}^{(h)}( w)}{n!}=\sum_{r+s=h\atop r,s\geqslant 0}\frac{ b_r\psi_{n,2}^{(s)}( w)}{r!s!}.
 \end{equation}
 Also, combining Eq.\,\eqref{eq:sSa} with $a^{(j)}( w)=0$, we can see that
 \begin{equation}\label{ls2}
  \sum_{k+l=j\atop k,l\geqslant 0}\frac{j!}{k!l!}\det\left(\varphi_{n,1}^{(k)}( w),\psi_{n,2}^{(l)}( w)\right)=0.
 \end{equation}
 Substituting Eq.\,\eqref{ls1} into Eq.\,\eqref{ls2} yields
 \begin{equation}
 \begin{split}
   0= &\det\left(\varphi_{n,1}^{(j)}( w),\psi_{n,2}( w)\right)+\sum_{\substack{l+r+s=j\\
  r+s\neq j\\
 l,r,s\geqslant 0}}\frac{j!}{l!r!s!}b_r\det\left(\psi_{n,2}^{(s)}( w),\psi_{n,2}^{(l)}( w)\right)\\
   =&\det\left(\varphi_{n,1}^{(j)}( w),\psi_{n,2}( w)\right)+\Big(\sum_{\substack{ l+r+s=j\\
  r+s\neq j\\
  l+r\neq j\\
 l,r,s\geqslant 0}}+\sum_{\substack{
  l+r+s=j\\
 r+s\neq j\\
  l+r= j\\
 l,r,s\geqslant 0
 }}
  \Big)\frac{j!}{l!r!s!}b_r\det\left(\psi_{n,2}^{(s)}( w),\psi_{n,2}^{(l)}( w)\right)\\
  =&\det\left(\varphi_{n,1}^{(j)}( w),\psi_{n,2}( w)\right)+\sum_{
  l+r=j\atop
 l>0,r\geqslant 0}\frac{j!}{l!r!}b_r\det\left(\psi_{n,2}( w),\psi_{n,2}^{(l)}( w)\right)\\
 =&\det\left(\varphi_{n,1}^{(j)}( w)-\sum_{
  l+r=j\atop
 l>0,r\geqslant 0}\frac{j!}{l!r!}b_r\psi_{n,2}^{(l)}( w),\psi_{n,2}( w)\right).
 \end{split}
 \end{equation}
 Since $\psi_{n,2}( w)$ is nonzero, there exists  a constant $b_j$ such that
 \begin{equation}
   \varphi_{n,1}^{(j)}( w)-\sum_{
  l+r=j\atop
 l>0,r\geqslant 0}\frac{j!}{l!r!}b_r\psi_{n,2}^{(l)}( w)=b_j\psi_{n,2}( w),
 \end{equation}
 i.e.,
 \begin{equation}
   \frac{\varphi_{n,1}^{(j)}( w)}{j!}=\sum_{
  l+r=j\atop
 l\geqslant 0,r\geqslant 0}\frac{b_r\psi_{n,2}^{(l)}( w)}{l!r!},
 \end{equation}
 which shows this proposition holds for $h=j$.
In summary, this proof has been established  by induction.
\end{proof}

\begin{corollary}\label{cormu}
If $ w$ is a zero of $a(\lambda)$ with multiplicity $m+1$, then for all $n\in\mathbb{Z}$ and each $h\in\{0,\ldots,m\}$,
\begin{equation}
  \frac{ \Phi_{n,1}^{(h)}( w)}{n!}=\sum_{j+k+l=h\atop j,k,l\geqslant 0} \frac{b_j\theta_n^{(k)}( w)\Psi_{n,2}^{(l)}(w)}{j!k!l!},
\end{equation}
where $\theta_n(\lambda)=\lambda^{-2n}$  and  $b_0,b_1,\ldots, b_m$ are given in Proposition \ref{prop:psi}.
\end{corollary}
\begin{proof}
It follows from Eq.\,\eqref{eq:mJost} and Proposition \ref{prop:psi} that
  \begin{equation}
  \begin{split}
      &\frac{ \Phi_{n,1}^{(h)}( w)}{h!}=\frac{(\theta_n^\frac12\varphi_{n,1})^{(h)}( w)}{h!}=\sum_{r+s=h\atop r,s\geqslant 0}
      \frac{(\theta_n^\frac12)^{(r)}( w)\varphi_{n,1}^{(s)}( w)}{r!s!}\\
      =&\sum_{r+s=h\atop r,s\geqslant 0}\sum_{j+m=s\atop j,m\geqslant 0}\frac{(\theta_n^\frac12)^{(r)}( w)b_j\psi_{n,2}^{(m)}( w)}{r!j!m!}
       =\sum_{r+j+m=h\atop r,j,m\geqslant 0}\frac{(\theta_n^\frac12)^{(r)}( w)b_j(\theta_n^\frac12\Psi_{n,2})^{(m)}( w)}{r!j!m!}\\
       =&\sum_{r+j+k+l=h\atop r,j,k,l\geqslant 0}\frac{(\theta_n^\frac12)^{(r)}( w)b_j(\theta_n^\frac12)^{(k)}( w)\Psi_{n,2}^{(l)}( w)}{r!j!k!l!}\\
       =&\sum_{j+k+l=h\atop j,k,l\geqslant 0}\sum_{r+s=k\atop r,s\geqslant 0}\frac{(\theta_n^\frac12)^{(r)}( w)(\theta_n^\frac12)^{(s)}( w)}{r!s!}\frac{b_j\Psi_{n,2}^{(l)}( w)}{j!l!}\\
      =&\sum_{j+k+l=h\atop j,k,l\geqslant 0} \frac{b_j\theta_n^{(k)}( w)\Psi_{n,2}^{(l)}( w)}{j!k!l!}.
  \end{split}
\end{equation}
\end{proof}
  Suppose that  $ w$ is a zero of $a(\lambda)$ with multiplicity $m+1$, then the transmission coefficient $\frac{1}{a(\lambda)}$ has a Laurent series expansion at $ \lambda= w$,
\begin{equation}
 \frac{1}{a(\lambda)}=\frac{a_{-m-1}}{( \lambda- w)^{m+1}}+\frac{a_{-m}}{( \lambda- w)^{m}}+\cdots+\frac{a_{-1}}{ \lambda- w}+O(1),\quad  \lambda\rightarrow w,
\end{equation}
where  $a_{-m-1}\neq 0$ and  $a_{-h-1}=\frac{\tilde{a}^{(m-h)}( w)}{(m-h)!}$, $\tilde{a}(\lambda)=\frac{( \lambda- w)^{m+1}}{a(\lambda)}$, $h=0,\ldots,m$.
Combining with Corollary \ref{cormu}, we obtain
 \begin{equation}
   \underset{ w}{\res}\frac{( \lambda- w)^h\Phi_{n,1}(\lambda)}{a(\lambda)}=\sum_{j+k+l+s=m-h\atop j,k,l,s\geqslant 0} \frac{\tilde{a}^{(j)}( w)b_k\theta_n^{(l)}(w)\Psi_{n,2}^{(s)}(w)}{j!k!l!s!},
 \end{equation}
 for each $h\in\{0,\ldots,m\}$.

Introduce a polynomial of degree at most $m$:
\begin{equation}\label{eq:f0}
  f_0(\lambda)=\sum_{h=0}^{m}\sum_{j+k=h\atop j,k\geqslant 0}\frac{\tilde{a}^{(j)}( w)b_k}{j!k!}( \lambda- w)^h,
\end{equation}
obviously, $f_0( w)\neq 0$. Therefore,
\begin{equation}\label{eq:resf0}
\begin{split}
  \underset{ w}{\res}\frac{( \lambda- w)^h\Phi_{n,1}(\lambda)}{a(\lambda)}=&\sum_{j+l+s=m-h\atop j,l,s\geqslant 0} \frac{f_0^{(j)}( w)\theta_n^{(l)}(w)\Psi_{n,2}^{(s)}( w)}{j!l!s!}\\
  =&\frac{[f_0(\lambda)\theta_n(\lambda)\Psi_{n,2}(\lambda)]^{(m-h)}|_{ \lambda= w}}{(m-h)!}.
\end{split}
 \end{equation}
   We refer to $f_0( w),\ldots,f_0^{(m)}( w)$ as the residue constants corresponding to  the discrete spectrum $ w$.

 \begin{assume}\label{sing}
    Suppose that $a(\lambda)$ has $N$  pairs of distinct zeros $ \pm\lambda_1,\ldots, \pm\lambda_N$ in $D_+$ with multiplicities $m_1+1,\ldots,m_N+1$, respectively. None of these zeros occur on $\Gamma$.
\end{assume}
\begin{lemma}\label{lemma:1}
Let $g(\lambda)$ be a complex-valued function, and suppose that
$w$ is an isolated singular point of $g(\lambda)$. Then, the following relations hold:
  \begin{equation}\label{eq:resrelation}
    \underset{w}{\res}g(\lambda)=-\underset{-w}{\res}g(-\lambda)=\overline{ \underset{\bar w}{\res}\overline{g(\bar\lambda)}}=-\underset{w^{-1}}{\res}\lambda^{-2}g(\lambda^{-1}).
  \end{equation}
  Let $h(\lambda)$  be a complex-valued function that is analytic at $\lambda_0$, then the $j$-th derivative of $h(\lambda)$ at  $\lambda_0$
  can be computed using the following formulas:
  \begin{equation}\label{eq:Tayrelation}
 \frac{\dd^j h(\lambda)}{\dd \lambda^j}|_{\lambda=\lambda_0}=\overline{\frac{\dd^j \overline{h(\bar\lambda)}}{\dd \lambda^j}|_{\lambda=\bar\lambda_0}}= (-1)^j\frac{\dd^j h(-\lambda)}{\dd \lambda^j}|_{\lambda=-\lambda_0}.
  \end{equation}
\end{lemma}
\begin{proof}
Suppose that $w$ is an $m$-th order pole of  $g(\lambda)$. Then, $g(\lambda)$ can be expanded into Laurent series in the neighborhood of $w$:
\begin{equation}
  g(\lambda)=\sum_{j=-m}^{+\infty}g_j(\lambda-w)^j,\quad \lambda\rightarrow w.
\end{equation}
 As a consequence,
 \begin{alignat}{2}
   &g(-\lambda)=\sum_{j=-m}^{+\infty}(-1)^jg_j(\lambda+w)^j,\quad &&-\lambda\rightarrow w,\\
   &\overline{g(\bar\lambda)}=\sum_{j=-m}^{+\infty}\bar{g}_j(\lambda-\bar w)^j,\quad& &\bar\lambda\rightarrow w.
    \end{alignat}
    As $\lambda^{-1}\rightarrow w$,
   \begin{equation}
      \begin{aligned}
         &\lambda^{-2}g(\lambda^{-1})=\lambda^{-2}\sum_{j=-m}^{+\infty}g_j(\lambda^{-1}-w)^j=\sum_{j=-m}^{+\infty}(-1)^jg_jw^j\lambda^{-(j+2)}(\lambda-w^{-1})^j\\
   =&\sum_{l=-m}^{+\infty}\sum_{j=-m}^{l}\frac{(-1)^jg_jw^j}{(l-j)!}\frac{\dd^{l-j}\left[\lambda^{-(j+2)}\right]}{\dd\lambda^{l-j}}|_{\lambda=w^{-1}}(\lambda-w^{-1})^{l}.
      \end{aligned}
    \end{equation}
From the above expansions, we find the following relations after calculating the corresponding residues:
 \begin{equation}
   \underset{w}{\res}g(\lambda)=-\underset{-w}{\res}g(-\lambda)=\overline{ \underset{\bar w}{\res}\overline{g(\bar\lambda)}}=-\underset{w^{-1}}{\res}\lambda^{-2}g(\lambda^{-1})=g_{-1}.
 \end{equation}
Now let us consider the Taylor series of $h(\lambda)$ in the neighborhood of $\lambda_0$:
 \begin{equation}
   h(\lambda)=\sum_{j=0}^{+\infty}\frac{\dd^j h(\lambda)}{\dd \lambda^j}|_{\lambda=\lambda_0}\frac{(\lambda-\lambda_0)^j}{j!},\quad \lambda\rightarrow\lambda_0.
 \end{equation}
 As a consequence,
 \begin{alignat}{2}
   &h(-\lambda)=\sum_{j=0}^{+\infty}\frac{\dd^j h(\lambda)}{\dd \lambda^j}|_{\lambda=\lambda_0}\frac{(-1)^j(\lambda+\lambda_0)^j}{j!},\quad &&-\lambda\rightarrow\lambda_0,\label{eq:hTay1}\\
   &\overline{h(\bar\lambda)}=\sum_{j=0}^{+\infty}\overline{\frac{\dd^j h(\lambda)}{\dd \lambda^j}|_{\lambda=\lambda_0}}\frac{(\lambda-\bar\lambda_0)^j}{j!},\quad& &\bar\lambda\rightarrow\lambda_0.\label{eq:hTay2}
 \end{alignat}
 Considering the Taylor series of $h(-\lambda)$ and $\overline{h(\bar\lambda)}$ at the neighborhood of $-\lambda_0$ and $\bar\lambda_0$, respectively, and comparing  the same power series with Eqs.\,\eqref{eq:hTay1} and \eqref{eq:hTay2}, we derive Eq.\,\eqref{eq:Tayrelation}.
\end{proof}
 \begin{prop}\label{prop:res}
   Suppose that $ \lambda_1,\ldots, \lambda_N$ satisfy the assumptions in Assumption \ref{sing}. there exists uniquely a  polynomial $f(\lambda)$ of degree less than $\displaystyle \mathcal{N}=N+\sum_{j=1}^N m_j$  with $f( \lambda_k)\neq 0$ such that, for  $k=1,\ldots,N$, $n_k=0,\ldots,m_k$,
   \begin{alignat}{2}
    &\underset{ \lambda_k}{\res}\frac{( \lambda- \lambda_k)^{n_k}\Phi_{n,1}(\lambda)}{a(\lambda)}=
      \frac{[f(\lambda)\theta_n(\lambda)\Psi_{n,2}(\lambda)]^{(m_k-n_k)}|_{ \lambda= \lambda_k}}{(m_k-n_k)!},\label{eq:resfa}\\
       &\underset{- \lambda_k}{\res}\frac{( \lambda+\lambda_k)^{n_k}\Phi_{n,1}(\lambda)}{a(\lambda)}=
      \frac{[f(-\lambda)\theta_n(-\lambda)\Psi_{n,2}(\lambda)]^{(m_k-n_k)}|_{ \lambda= -\lambda_k}}{(-1)^{m_k}(m_k-n_k)!},\label{eq:resfaa}\\
           &\underset{ \bar\lambda_k}{\res}\frac{( \lambda- \bar\lambda_k)^{n_k}\Phi_{n,2}(\lambda^{-1})}{\overline{a( \bar{\lambda})}}=
    -\frac{[\overline{f(\bar\lambda)}\overline{\theta_n(\bar\lambda)}\Psi_{n,1}(\lambda^{-1})]^{(m_k-n_k)}|_{ \lambda= \bar\lambda_k}}{(m_k-n_k)!},\label{eq:resfb}\\
      &\underset{ -\bar\lambda_k}{\res}\frac{( \lambda+\bar\lambda_k)^{n_k}\Phi_{n,2}(\lambda^{-1})}{\overline{a( \bar{\lambda})}}=
    \frac{[\overline{f(-\bar\lambda)}\overline{\theta_n(-\bar\lambda)}\Psi_{n,1}(\lambda^{-1})]^{(m_k-n_k)}|_{ \lambda= -\bar\lambda_k}}{(-1)^{m_k+1}(m_k-n_k)!}.\label{eq:resfbb}
   \end{alignat}
 \end{prop}
 \begin{proof}
  Similar to Eq.\,\eqref{eq:resf0}, for each $k\in\{1,\ldots, N\}$, there exists a polynomial $f_k(\lambda)$ of degree at most $m_k$  with $f_k( \lambda_k)\neq 0$ such that
 \begin{equation}
   \underset{ \lambda_k}{\res}\frac{( \lambda- \lambda_k)^{n_k}\Phi_{n,1}(\lambda)}{a(\lambda)}=
      \frac{[f_k(\lambda)\theta_n(\lambda)\Psi_{n,2}(\lambda)]^{(m_k-n_k)}|_{ \lambda= \lambda_k}}{(m_k-n_k)!}, \quad n_k=0,\ldots,m_k.
 \end{equation}
   Using the Hermite interpolation formula, there exists uniquely a  polynomial $f(\lambda)$ of degree less than $\displaystyle \mathcal{N}$ satisfying
    \begin{equation}\label{eq:feqn}
    \begin{cases}
      f^{(n_1)}( \lambda_1)=f_1^{(n_1)}( \lambda_1), \quad& n_1=0,\ldots,m_1,\\
     \qquad\qquad \vdots&\\
       f^{(n_N)}( \lambda_N)=f_N^{(n_N)}( \lambda_N), \quad& n_N=0,\ldots,m_N.
    \end{cases}
    \end{equation}
    We have shown Eq.\,\eqref{eq:resfa}.  Due to the  symmetry \eqref{eq:Jostsym1} and Lemma \ref{lemma:1},
    \begin{equation}
    \begin{split}
            &\underset{ \bar\lambda_k}{\res}\frac{( \lambda-\bar\lambda_k)^{n_k}\Phi_{n,2}( \lambda^{-1})}{\overline{a( \bar\lambda)}}
            =\underset{ \bar\lambda_k}{\res}\frac{( \lambda- \bar\lambda_k)^{n_k}(-\ii \sigma_2)\overline{\Phi_{n,1}(\bar\lambda)}}{\overline{a(\bar\lambda)}}\\
            =&-\ii \sigma_2\overline{\underset{ \lambda_k}{\res}\frac{( \lambda- \lambda_k)^{n_k}\Phi_{n,1}(\lambda)}{a(\lambda)}}=-\ii\sigma_2\overline{\frac{[f(\lambda)\theta_n(\lambda)\Psi_{n,2}(\lambda)]^{(m_k-n_k)}|_{ \lambda= \lambda_k}}{(m_k-n_k)!}}\\
             =& -\ii \sigma_2\frac{[\overline{f(\bar\lambda)\theta_n(\bar\lambda)\Psi_{n,2}(\bar\lambda)}]^{(m_k-n_k)}|_{ \lambda= \bar\lambda_k}}{(m_k-n_k)!}= -\frac{[\overline{f(\bar\lambda)\theta_n(\bar\lambda)}\Psi_{n,1}(\lambda^{-1})]^{(m_k-n_k)}|_{ \lambda= \bar\lambda_k}}{(m_k-n_k)!}.
    \end{split}
    \end{equation}
    We have shown Eq.\,\eqref{eq:resfb}. Furthermore, due to the symmetry  \eqref{eq:Jostsym2} and Lemma \ref{lemma:1},
     \begin{equation}
       \begin{split}
        & \underset{- \lambda_k}{\res}\frac{( \lambda+\lambda_k)^{n_k}\Phi_{n,1}(\lambda)}{a(\lambda)}=(-1)^{n_k}\sigma_3\underset{- \lambda_k}{\res}\frac{( -\lambda-\lambda_k)^{n_k}\Phi_{n,1}(-\lambda)}{a(-\lambda)}\\
        =&(-1)^{n_k+1}\sigma_3\underset{ \lambda_k}{\res}\frac{( \lambda-\lambda_k)^{n_k}\Phi_{n,1}(\lambda)}{a(\lambda)}=\sigma_3\frac{[f(\lambda)\theta_n(\lambda)\Psi_{n,2}(\lambda)]^{(m_k-n_k)}|_{ \lambda= \lambda_k}}{(-1)^{n_k+1}(m_k-n_k)!}\\
        =&\frac{[f(\lambda)\theta_n(\lambda)\Psi_{n,2}(-\lambda)]^{(m_k-n_k)}|_{ \lambda= \lambda_k}}{(-1)^{n_k}(m_k-n_k)!}   =\frac{[f(-\lambda)\theta_n(-\lambda)\Psi_{n,2}(\lambda)]^{(m_k-n_k)}|_{ \lambda= -\lambda_k}}{(-1)^{m_k}(m_k-n_k)!}.
       \end{split}
     \end{equation}
   We have shown Eq.\,\eqref{eq:resfaa}.  Similarly, we can obtain Eq.\,\eqref{eq:resfbb}.
 \end{proof}

\section{Inverse scattering problem}\label{inverse}
The RH problem is a key component in formulating the inverse scattering problem, allowing for the establishment of a jump condition that relates the meromorphic eigenfunctions in the inside and outside of unit disk.
\subsection{ Stationary Riemann--Hilbert problem}
We consider the  $2\times 2$ matrix-valued function $\mathbf{M}_n(\lambda)$ defined by
  \begin{equation}\label{Mdef}
\mathbf{M}_n(\lambda):=\begin{cases}
  \begin{pmatrix}
                              1 & 0 \\
                              0 & \delta_n^+
                            \end{pmatrix}\left(\frac{\Phi_{n,1}(\lambda)}{a(\lambda)},\Psi_{n,2}(\lambda)\right),\quad &\lambda\in D_+,\\[3ex]
                            \begin{pmatrix}
                              1 & 0 \\
                              0 & \delta_n^+
                            \end{pmatrix}\left(\Psi_{n,1}(\lambda),\frac{\Phi_{n,2}(\lambda)}{\overline{a(\bar \lambda^{-1})}}\right),\quad &\lambda\in D_-.
\end{cases}
 \end{equation}
  \begin{figure}[!htb]
\centering
\begin{tikzpicture}[>=stealth,scale=0.5]
  \fill[gray!50] (0,0) -- (0:3) arc (0:360:3) -- cycle;
 \draw[->,fill=black](5,0)--(5.5,0)node[right]{\tiny$\operatorname{Re} \lambda$};
\draw[->](0,4.95)--(0,5.5)node[above]{\tiny$\operatorname{Im} \lambda$};
  \draw[->](271:3)arc(271:90:3)node[above]{\tiny$\Gamma$};
  \draw[->](360:3)arc(360:270:3);
  \draw(0:3)arc(0:91:3);
  \node at (45:4.5) {\tiny$D_+$};
  \node at (0:0) {\tiny$D_-$};
 \draw[dashed](225:7) rectangle(45:7);
\end{tikzpicture}
\caption{ The regions $D_+$ (white), $D_-$ (shaded) and the oriented contour $\Gamma$.}
\label{fig1}
\end{figure}
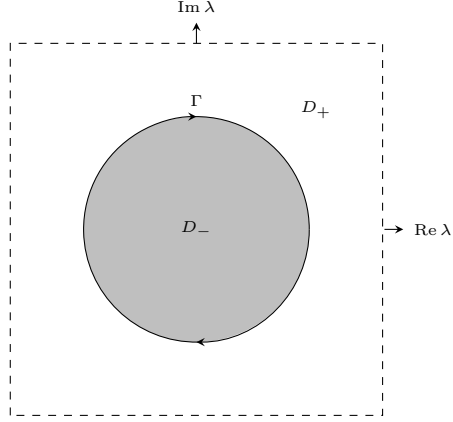
 From this, we can infer that $\mathbf{M}_n(\lambda)$ satisfying the following properties:
\begin{itemize}
  \item \textbf{Analyticity:} $\mathbf{M}_n(\lambda)$ is an analytic function of $\lambda$ for $\lambda\in\mathbb{C}\backslash\left(\Gamma\cup\{\pm\lambda_k,\pm\bar{\lambda}^{-1}_k\}_{k=1}^N\right)$;
  \item \textbf{Residues:} For $ k=1,\ldots,N$, at $\lambda=\pm\lambda_k$ and $\lambda=\pm\bar{\lambda}_k$, $\mathbf{M}_n(\lambda)$ and $\mathbf{M}_n(\lambda^{-1})$ have  multiple-poles of order $m_k+1$, respectively, and their residues satisfy the following conditions:
\begin{subequations}\label{eq:res1}
  \begin{align}
    &\underset{\lambda_k}{\res} (\lambda-\lambda_k)^{n_k}\mathbf{M}_n(\lambda)=\left(\frac{[f(\lambda) \theta_n(\lambda)\mathbf{M}_{n,2}(\lambda)]^{(m_k-n_k)}|_{\lambda=\lambda_k}}{(m_k-n_k)!},\mathbf{0}\right),  \label{eq:res11}\\
      &\underset{-\lambda_k}{\res} (\lambda+\lambda_k)^{n_k}\mathbf{M}_n(\lambda)=\left(\frac{[f(-\lambda) \theta_n(-\lambda)\mathbf{M}_{n,2}(\lambda)]^{(m_k-n_k)}|_{\lambda=-\lambda_k}}{(-1)^{m_k}(m_k-n_k)!},\mathbf{0}\right),  \label{eq:res111}\\
    &\underset{\bar\lambda_k}{\res} (\lambda-\bar\lambda_k)^{n_k}\mathbf{M}_n(\lambda^{-1})=\left(\mathbf{0},-\frac{[\overline{f(\bar\lambda) \theta_n(\bar\lambda)}\mathbf{M}_{n,1}(\lambda^{-1})]^{(m_k-n_k)}|_{\lambda=\bar\lambda_k}}{(m_k-n_k)!}\right),\label{eq:res12}\\
    &\underset{-\bar\lambda_k}{\res} (\lambda+\bar\lambda_k)^{n_k}\mathbf{M}_n(\lambda^{-1})=\left(\mathbf{0},\frac{[\overline{f(-\bar\lambda) \theta_n(-\bar\lambda)}\mathbf{M}_{n,1}(\lambda^{-1})]^{(m_k-n_k)}|_{\lambda=-\bar\lambda_k}}{(-1)^{m_k+1}(m_k-n_k)!}\right),\label{eq:res112}
  \end{align}
\end{subequations}
for  $n_k=0,\ldots, m_k$;
  \item \textbf{Jump:} $\mathbf{M}_n(\lambda)$ has a jump across the oriented contour $\Gamma$ as in figure \ref{fig1}:
  \begin{equation}\label{Mjump}
  \mathbf{M}_{n+}(\lambda)=\mathbf{M}_{n-}(\lambda)\mathbf{J}_n(\lambda),\quad \lambda\in\Gamma,
\end{equation}
where $ \mathbf{M}_{n\pm}(\lambda)=\displaystyle\overset{\angle}{\lim_{\substack{\lambda'\rightarrow \lambda\\ \lambda'\in D_{\pm}}}}\mathbf{M}_n(\lambda')$, `` $\overset{\angle}{\lim}$" means non-tangential  limit,   and the jump matrix reads
\begin{equation}\label{eq:jump}
  \mathbf{J}_n(\lambda)=\begin{pmatrix}
    1+\overline{\gamma(\bar \lambda^{-1})}\gamma(\lambda)& \theta^{-1}_n(\lambda)\overline{\gamma(\bar\lambda^{-1})}\\
     \theta_n(\lambda)\gamma(\lambda)&1
  \end{pmatrix};
\end{equation}
  \item \textbf{Normalizations:} $\mathbf{M}_n(\lambda)$ has the following asymptotics:
\begin{subequations}\label{eq:masy}
\begin{alignat}{2}
&\mathbf{M}_n(\lambda)=\mathbf{I}+O(\lambda^{-1}),\qquad &&\lambda\rightarrow\infty,\\
&\mathbf{M}_n(\lambda)=\begin{pmatrix}
                           (\delta^+_n)^{-1} & 0 \\
                           0 & \delta_n^+
                         \end{pmatrix}+O(\lambda),\qquad &&\lambda\rightarrow 0.
\end{alignat}
\end{subequations}
\end{itemize}

In Sect.\,\ref{sec:dir}, we  have established the direct scattering map from the initial data to the scattering data:
\begin{equation}
  \mathcal{D}: q_n\mapsto \left\{\gamma(\lambda),\left\{\lambda_k, f(\lambda_k), f^{(1)}(\lambda_k),\ldots, f^{(m_k)}(\lambda_k) \right\}_{k=1}^N\right\}.
\end{equation}
Indeed, the matrix $\mathbf{M}_n(\lambda)$ can be  recovered from the scattering data by solving a $2\times 2$ matrix
RH problem, and then the potential  $q_n$ can be reconstructed in terms of $\mathbf{M}_n(\lambda)$.
\subsection{Time evolution and reconstruction }

If the potential  $q_n$ is replaced by a time-dependent potential  $q_n(t)$ that satisfies the  Ablowitz--Ladik equation \eqref{eq:dfnls}, then the scattering data become time dependent as well:
\begin{equation}
 \gamma(\lambda;t),\left\{\lambda_k(t), f(\lambda_k;t), f^{(1)}(\lambda_k;t),\ldots, f^{(m_k)}(\lambda_k;t)\right\}_{k=1}^N.
\end{equation}
Indeed, the discrete spectrum $\lambda_k(t)=\lambda_k$ does not change in time, the reflection coefficient and residue constants evolves in time as follows:
\begin{equation}
  \gamma(\lambda;t)=\gamma(\lambda)\e^{\ii(\lambda-\lambda^{-1})^2t},\quad f^{(n_k)}(\lambda_k;t)=\partial^{n_k}_\lambda[f(\lambda)\e^{\ii(\lambda-\lambda^{-1})^2t}]|_{\lambda=\lambda_k}.
\end{equation}
In the following, we will consider the inverse scattering map
\begin{equation}
  \mathcal{I}: \left\{\gamma(\lambda;t),\left\{\lambda_k, f(\lambda_k;t), f^{(1)}(\lambda_k;t),\ldots, f^{(m_k)}(\lambda_k;t)\right\}_{k=1}^N\right\}\mapsto q_n(t),
\end{equation}
which can be characterised in terms of a $2\times 2$ matrix RH problem.
\begin{problem}\label{frhp}
  Seek a $2\times 2$ matrix-valued analytic function $\mathbf{M}_n(\lambda;t)$ except for the contour $\Gamma$ and the finite point set $\{\pm\lambda_k,\pm\bar{\lambda}^{-1}_k\}_{k=1}^N$, satisfying  the same residue conditions, jump relation and asymptotic behaviours  as in Eqs.\,\eqref{eq:res1}-\eqref{eq:masy}  but with $f^{(n_k)}(\lambda_k)$  and $\gamma(\lambda)$ replaced by $f^{(n_k)}(\lambda_k;t)$ and $\gamma(\lambda;t)$, respectively.
\end{problem}

For each $k$, let $\Omega_{ k}$ be a small disk centered at $\lambda_k$ with sufficiently small radius such that it lies in the domain $D_+$ and is disjoint from all other disks and $\{\Omega_{-k}\}_{k=1}^N$, where $\Omega_{-k}=\{\lambda\in D_+|-\lambda\in \Omega_k\}$.  Additionally, let $\tilde{\Omega}_{\pm k}=\{\lambda\in D_-|\pm\bar\lambda^{-1}\in \Omega_k\}$. We define  a new matrix unknown $\tilde{\mathbf{M}}_n(\lambda;t)$ in terms of $\mathbf{M}_n(\lambda;t)$ by
\begin{equation}\label{eq:tMdef}
  \tilde{\mathbf{M}}_n(\lambda;t)=\begin{cases}
    \mathbf{M}_n(\lambda;t)\mathbf{P}_{n}(k,\lambda;t),\quad &\lambda\in\Omega_k,\quad k=1,\ldots,N,\\
    \mathbf{M}_n(\lambda;t)\mathbf{P}_{n}^{-1}(k,-\lambda;t),\quad &\lambda\in\Omega_{-k},\quad k=1,\ldots,N,\\
     \mathbf{M}_n(\lambda;t)[\mathbf{P}^\dag_{n}(k,\bar\lambda^{-1};t)]^{-1},\quad &\lambda\in\tilde\Omega_k,\quad k=1,\ldots,N,\\
    \mathbf{M}_n(\lambda;t)\mathbf{P}^\dag_{n}(k,-\bar\lambda^{-1};t),\quad &\lambda\in\tilde\Omega_{-k},\quad k=1,\ldots,N,\\
    \mathbf{M}_n(\lambda;t),\quad &\text{otherwise},
  \end{cases}
\end{equation}
where \begin{equation}\label{eq:Pndef}
  \mathbf{P}_{n}(k,\lambda;t)=\begin{pmatrix}
      1&0\\
      -\frac{f(\lambda;t)\theta_n(\lambda)}{(\lambda-\lambda_k)^{m_k+1}}&1
    \end{pmatrix},
\end{equation} ``\dag'' denotes the  conjugate transpose. It can be observed that the matrix $\mathbf{M}_n(\lambda;t)\mathbf{P}_{n}(k,\lambda;t)$ has a removeable singularity at $\lambda_k$.  Indeed,
    \begin{equation}
    \begin{split}
       &\underset{\lambda_k}{\res} (\lambda-\lambda_k)^{n_k}\tilde{\mathbf{M}}_n(\lambda;t)\\
       =&\underset{\lambda_k}{\res}(\lambda-\lambda_k)^{n_k}\left( \mathbf{M}_{n,1}(\lambda;t)-\frac{f(\lambda;t)\theta_n(\lambda)}{(\lambda-\lambda_k)^{m_k+1}}\mathbf{M}_{n,2}(\lambda;t), \mathbf{M}_{n,2}(\lambda;t)\right)\\
       =&\left(\underset{\lambda_k}{\res}(\lambda-\lambda_k)^{n_k}\mathbf{M}_{n,1}(\lambda;t)-\underset{\lambda_k}{\res}\frac{f(\lambda;t)\theta_n(\lambda)\mathbf{M}_{n,2}(\lambda;t)}{(\lambda-\lambda_k)^{m_k-n_k+1}},\mathbf{0}\right),
    \end{split}
    \end{equation}
     this  can be seen by considering the residue condition  \eqref{eq:res11} and considering the Taylor series of $f(\lambda;t)\theta_n(\lambda)\mathbf{M}_{n,2}(\lambda;t)$ at $\lambda_k$. It follows that for each $k$ and any $0\leqslant n_k\leqslant m_k$, we have $ \underset{\lambda_k}{\res} (\lambda-\lambda_k)^{n_k}\tilde{\mathbf{M}}_n(\lambda;t)=\mathbf{0}$. By completely analogous reasoning, it can be shown that the matrix $ \mathbf{M}_n(\lambda;t)\mathbf{P}_{n}^{-1}(k,-\lambda;t)$  has a removable singularity
at $-\lambda_k$, the matrix $ \mathbf{M}_n(\lambda^{-1};t)[\mathbf{P}^\dag_{n}(k,\bar\lambda;t)]^{-1}$  has a removable singularity
at $\bar\lambda_k$, the matrix $ \mathbf{M}_n(\lambda^{-1};t)\mathbf{P}^\dag_{n}(k,-\bar\lambda;t)$  has a removable singularity
at $-\bar\lambda_k$. Based on the definition \eqref{eq:tMdef} of $\tilde{\mathbf{M}}_n(\lambda;t)$, it can be concluded that all of $\{\pm\lambda_k,\pm\bar\lambda_k^{-1}\}_{k=1}^N$ are the removable singularities of $\tilde{\mathbf{M}}_n(\lambda;t)$.
Therefore, it can be seen that  $\tilde{\mathbf{M}}_n(\lambda;t)$ satisfies the conditions of an equivalent RH problem closely related
to RH Problem \ref{frhp} but with the residue conditions replaced by jump conditions across small circles centered at the points $\{\pm\lambda_k,\pm\bar\lambda^{-1}_k\}_{k=1}^N$.
    \begin{problem}\label{frhptm}
  Seek a $2\times 2$ matrix-valued   function $\tilde{\mathbf{M}}_n(\lambda;t)$ satisfying the following properties:
  \begin{itemize}
  \item \textbf{Analyticity:} $\tilde{\mathbf{M}}_n(\lambda;t)$ is an analytic function of $\lambda$ for $\lambda\in\mathbb{C}\backslash\Sigma$ where $\Sigma=\Gamma\cup\{\partial\Omega_{\pm k},\partial\tilde{\Omega}_{\pm k}\}_{k=1}^N$;
  \item \textbf{Jump:} The matrix $\tilde{\mathbf{M}}_n(\lambda;t)$ takes continuous boundary values $\tilde{\mathbf{M}}_{n\pm}(\lambda;t)$ on $\Gamma$ from $D_\pm$, as well as from the left and right on $\partial\Omega_{\pm1},\ldots,\partial\Omega_{\pm N}$ oriented  in a clockwise direction and $\partial\tilde\Omega_{\pm1},\ldots,\partial\tilde\Omega_{\pm N}$ oriented  in a counterclockwise direction. The boundary values are related
  \begin{equation}\label{tMjump}
  \tilde{\mathbf{M}}_{n+}(\lambda;t)=\tilde{\mathbf{M}}_{n-}(\lambda;t)\tilde{\mathbf{J}}_n(\lambda;t),\quad \lambda\in\Sigma,
\end{equation}
where
\begin{equation}\label{eq:tjump}
  \tilde{\mathbf{J}}_n(\lambda;t)=\begin{cases}\mathbf{J}_n(\lambda;t),\quad &\lambda\in\Gamma,\\
  \mathbf{P}_n^{-1}(k,\lambda;t),\quad &\lambda\in\partial\Omega_k,\quad k=1,\ldots,N,\\
  \mathbf{P}_n(k,-\lambda;t),\quad &\lambda\in\partial\Omega_{-k},\quad k=1,\ldots,N,\\
  [\mathbf{P}_n^{\dag}(k,\bar\lambda^{-1};t)]^{-1},\quad &\lambda\in\partial\tilde\Omega_k,\quad k=1,\ldots,N,\\
  \mathbf{P}_n^\dag(k,-\bar\lambda^{-1};t),\quad &\lambda\in\partial\tilde\Omega_{-k},\quad k=1,\ldots,N,
  \end{cases}
\end{equation}
with $\mathbf{J}_n(\lambda;t)=
    \begin{pmatrix}
    1+\overline{\gamma(\bar \lambda^{-1};t)}\gamma(\lambda;t)& \theta^{-1}_n(\lambda)\overline{\gamma(\bar\lambda^{-1};t)}\\
     \theta_n(\lambda)\gamma(\lambda;t)&1
  \end{pmatrix}$ and $\mathbf{P}_n(k,\lambda;t) $ is defined in Eq.\,\eqref{eq:tMdef};
  \item \textbf{Normalizations:} $\tilde{\mathbf{M}}_n(\lambda;t)$ has the following asymptotics:
\begin{subequations}\label{eq:tmasy}
\begin{alignat}{2}
&\tilde{\mathbf{M}}_n(\lambda;t)=\mathbf{I}+O(\lambda^{-1}),\qquad &&\lambda\rightarrow\infty,\\
&\tilde{\mathbf{M}}_n(\lambda;t)=\begin{pmatrix}
                           (\delta^+_n)^{-1} & 0 \\
                           0 & \delta_n^+
                         \end{pmatrix}+O(\lambda),\qquad &&\lambda\rightarrow 0.
\end{alignat}
\end{subequations}
\end{itemize}
\end{problem}
The solution of the pole-free RH problem \ref{frhptm} exists uniquely if and only if the following lemma holds (see Theorem 9.3 in Ref.\,\cite{Zhou1989}).
\begin{lemma}{(Vanishing Lemma)}\label{vanish}
The  RH problem for $\tilde{\mathbf{M}}_n(\lambda;t)$ obtained from   RH problem \ref{frhptm} by replacing the asymptotics \eqref{eq:tmasy} by
\begin{alignat*}{2}
 &\tilde{\mathbf{M}}_n(\lambda;t)=O(\lambda^{-1}), \qquad&& \lambda\rightarrow\infty,\\
  & \tilde{\mathbf{M}}_n(\lambda;t)= O(\lambda), \quad&& \lambda\rightarrow 0,
\end{alignat*}
 has only the zero solution.
\end{lemma}
\begin{proof}
 Let \begin{equation}
   \mathcal{H}_n(\lambda;t)=\tilde{\mathbf{M}}_n(\lambda;t)\tilde{\mathbf{M}}^\dag_n(\bar\lambda^{-1};t),
 \end{equation} then
$\mathcal{H}_n(\lambda;t)$ is an analytic function of $\lambda$ for $\lambda\in\mathbb{C}\backslash\Sigma$, and  $\mathcal{H}_n(\lambda;t)$ is also continuous up to $\Sigma$. We calculate the jump of $\mathcal{H}_n(\lambda;t)$ for  $\lambda\in\Sigma$ as follows: first observe that   \begin{equation}
  \mathcal{H}_{n+}(\lambda;t)=\tilde{\mathbf{M}}_{n+}(\lambda;t)\tilde{\mathbf{M}}_{n-}^\dag(\bar\lambda^{-1};t),
\end{equation}
for example, if $\lambda\rightarrow\partial\Omega_k$ from the left (``$+$"  side), then $\bar\lambda^{-1}\rightarrow\partial\tilde\Omega_k$
from the right (``$-$"  side), where the ``$\pm$" subscripts indicate boundary values on $\partial\Omega_k$ (for $\lambda$) and $\partial\tilde{\Omega}_k$ (for $\bar\lambda^{-1}$). Applying the jump conditions across $\Sigma$ gives
\begin{equation}\label{eq:Hp}
  \mathcal{H}_{n+}(\lambda;t)=\tilde{\mathbf{M}}_{n+}(\lambda;t)\tilde{\mathbf{M}}_{n-}^\dag(\bar\lambda^{-1};t)=\tilde{\mathbf{M}}_{n-}(\lambda;t)\tilde{\mathbf{J}}_n(\lambda;t)\tilde{\mathbf{M}}_{n-}^\dag(\bar\lambda^{-1};t).
\end{equation}
Using the relation $\tilde{\mathbf{J}}_n(\lambda;t)=\tilde{\mathbf{J}}^\dag_n(\bar\lambda^{-1};t)$, we find
\begin{equation}
  \mathcal{H}_{n+}(\lambda;t)=\tilde{\mathbf{M}}_{n-}(\lambda;t)\tilde{\mathbf{J}}^\dag_n(\bar\lambda^{-1};t)\tilde{\mathbf{M}}_{n-}^\dag(\bar\lambda^{-1};t)=\tilde{\mathbf{M}}_{n-}(\lambda;t)\tilde{\mathbf{M}}_{n+}^\dag(\bar\lambda^{-1};t)=\mathcal{H}_{n-}(\lambda;t).
\end{equation}
 which imply $\mathcal{H}_n(\lambda;t)$ is continuous in the whole complex $\lambda$-plane. It then follows by Morera's Theorem that $\mathcal{H}_n(\lambda;t)$ is an entire function of $\lambda$.
Observing that $\lim_{\lambda\rightarrow\infty}\mathcal{H}_n(\lambda;t)=\mathbf{0}$,
  by  Liouville's theorem, we get
\begin{equation}
  \mathcal{H}_n(\lambda;t)\equiv\mathbf{0}.
\end{equation}
It follows from that
$\tilde{\mathbf{J}}_n(\lambda;t)$ is positive definite for $\lambda\in\Gamma$, combining with Eq.\,\eqref{eq:Hp} yields  $\tilde{\mathbf{M}}_{n-}(\lambda;t)=\mathbf{0}$ for $\lambda\in \Gamma$.
 By the jump condition, we also get $\tilde{\mathbf{M}}_{n+}(\lambda;t)=\mathbf{0}$ for $\lambda\in\Gamma$. It follows by analytic continuation that $\tilde{\mathbf{M}}_n(\lambda;t)=\mathbf{0}$ holds as an identity all the way up to $\{\partial\Omega_{\pm k},\partial\tilde{\Omega}_{\pm k}\}_{k=1}^N$. But then applying the jump condition for $\tilde{\mathbf{M}}_n(\lambda;t)$ on these arcs shows that again $\tilde{\mathbf{M}}_n(\lambda;t)=\mathbf{0}$ for $\lambda$ lies in the interior of  $\{\Omega_{\pm k},\tilde{\Omega}_{\pm k}\}_{k=1}^N$. Thus $\tilde{\mathbf{M}}_n(\lambda;t)=\mathbf{0}$ on the whole complex plane as desired.
\end{proof}
\begin{theorem}\label{Thinv}
  If $\tilde{\mathbf{M}}_n(\lambda;t)$ is the solution of RH problem \ref{frhptm}, then
  \begin{equation}\label{eq:recon1}
      q_n(t)= \lim_{\lambda\rightarrow 0}\frac{\tilde{\mathbf{M}}_{n+1,12}(\lambda;t)}{\lambda},
  \end{equation}
 solves the  Ablowitz--Ladik equation \eqref{eq:dfnls}.
\end{theorem}
\begin{proof}
From the symmtries
\begin{equation}
  \tilde{\mathbf{J}}_n(\lambda;t)=\sigma_2 \overline{\tilde{\mathbf{J}}_n^{-1}(\bar\lambda^{-1};t)}\sigma_2,\quad \tilde{\mathbf{J}}_n(\lambda;t)=\sigma_3 \tilde{\mathbf{J}}_n(-\lambda;t)\sigma_3,
\end{equation}
and the asymptotics in RH problem \ref{frhptm}, it follows  that  $\tilde{\mathbf{M}}_n(\lambda;t)$, $\begin{pmatrix}
                           (\delta^+_n)^{-1} & 0 \\
                           0 & \delta_n^+
                         \end{pmatrix}\sigma_2\overline{\tilde{\mathbf{M}}_n(\bar\lambda^{-1};t)}\sigma_2$ and  $\tilde{\mathbf{M}}_n(\lambda;t)=\sigma_3\tilde{\mathbf{M}}_n(-\lambda;t)\sigma_3$ satisfy the same RH problem. Based on the uniqueness, we conclude
\begin{equation}\label{eq:tMsym}
  \tilde{\mathbf{M}}_n(\lambda;t)=\begin{pmatrix}
                           (\delta^+_n)^{-1} & 0 \\
                           0 & \delta_n^+
                         \end{pmatrix}\sigma_2\overline{\tilde{\mathbf{M}}_n(\bar\lambda^{-1};t)}\sigma_2,\quad  \tilde{\mathbf{M}}_n(\lambda;t)=\sigma_3\tilde{\mathbf{M}}_n(-\lambda;t)\sigma_3.
\end{equation}
 Let
 \begin{equation}\label{retmhm}
   \hat{\mathbf{M}}_n(\lambda;t)=\begin{pmatrix}
     \delta_n^+&0\\0&1
   \end{pmatrix}\tilde{\mathbf{M}}_n(\lambda;t),
 \end{equation}
 it follows from the symmetries in Eq.\,\eqref{eq:tMsym} that
 \begin{equation}\label{eq:hMsym}
   \hat{\mathbf{M}}_n(\lambda;t)= \sigma_2\overline{\hat{\mathbf{M}}_n(\bar\lambda^{-1};t)}\sigma_2,\quad  \hat{\mathbf{M}}_n(\lambda;t)=\sigma_3\hat{\mathbf{M}}_n(-\lambda;t)\sigma_3.
 \end{equation}
According to the asymptotics of $\tilde{\mathbf{M}}_n(\lambda;t)$ for $\lambda$ approaches infinity and zero, we assume that $\hat{\mathbf{M}}_n(\lambda;t)$ has the asymptotic expansion forms
   \begin{alignat}{2}
    &\hat{\mathbf{M}}_n(\lambda;t)=\sum_{k=0}^\infty\frac{\hat{\mathbf{M}}_n^{(k)}(t)}{\lambda^k},\qquad &&\lambda\rightarrow\infty,\label{tMasy1}\\
    &\hat{\mathbf{M}}_n(\lambda;t)=\sum_{k=0}^\infty\mathcal{M}_n^{(k)}(t)\lambda^k,\qquad &&\lambda\rightarrow 0,
  \end{alignat}
  where \begin{equation}
    \hat{\mathbf{M}}_n^{(0)}(t)=\begin{pmatrix}
                           \delta_n^+ & 0 \\
                           0 & 1
                         \end{pmatrix},\quad \mathcal{M}_n^{(0)}(t)=\begin{pmatrix}
                           1 & 0 \\
                           0 &\delta_n^+
                         \end{pmatrix}.
  \end{equation}
  Taking
into account the symmetries in Eq.\,\eqref{eq:hMsym}, we find that   for any positive integer $k$, $\hat{\mathbf{M}}_n^{(2k-1)}(t)$ and $\mathcal{M}_n^{(2k-1)}(t)$ are off-diagonal, $\hat{\mathbf{M}}_n^{(2k)}(t)$ and $\mathcal{M}_n^{(2k)}(t)$ are  diagonal,
\begin{equation}\label{sym:hmcalm}
  \hat{\mathbf{M}}_n^{(k)}(t)=\sigma_2\overline{\mathcal{M}_n^{(k)}(t)}\sigma_2.
\end{equation}
Let
\begin{alignat}{1}
   & \mathscr{A}_n(\lambda;t)=\hat{\mathbf{M}}_{n}(\lambda;t)-\Lambda^{-1}\hat{\mathbf{M}}_{n+1}(\lambda;t)\Lambda+\mathbf{Q}_n(t)\hat{\mathbf{M}}_{n+1}(\lambda;t)\Lambda,\label{eq:mscra}\\
   & \mathscr{B}_n(\lambda;t)=\partial_t\hat{\mathbf{M}}_n(\lambda;t)-\mathbf{V}_n(\lambda;t)\hat{\mathbf{M}}_n(\lambda;t)-\frac{\ii}{2}(\lambda-\lambda^{-1})^2\hat{\mathbf{M}}_n(\lambda;t)\sigma_3,
\end{alignat}
  where $\mathbf{Q}_n(t)$ and $\mathbf{V}_n(\lambda;t)$ are defined by Eq.\,\eqref{uvdef}, then $\mathscr{A}_{n}(\lambda;t)$ and $\mathscr{B}_{n}(\lambda;t)$ satisfy the same jump condition as in RH problem \ref{frhptm}, i.e.,
  \begin{equation}
    \mathscr{A}_{n+}(\lambda;t)=\mathscr{A}_{n-}(\lambda;t)\tilde{\mathbf{J}}_{n}(\lambda;t),\quad \mathscr{B}_{n+}(\lambda;t)=\mathscr{B}_{n-}(\lambda;t)\tilde{\mathbf{J}}_{n}(\lambda;t),\quad \lambda\in\Sigma.
  \end{equation}
  In addition, it follows from the first symmetry in Eq.\,\eqref{eq:hMsym} that
  \begin{equation}\label{sym:mscrab}
    \mathscr{A}_n(\lambda;t)= \sigma_2\overline{\mathscr{A}_n(\bar\lambda^{-1};t)}\sigma_2, \quad     \mathscr{B}_n(\lambda;t)= \sigma_2\overline{\mathscr{B}_n(\bar\lambda^{-1};t)}\sigma_2.
  \end{equation}
  If $q_n(t)$ is defined by Eq.\,\eqref{eq:recon1}, then the relation \eqref{retmhm} and the symmetry \eqref{sym:hmcalm} imply
  \begin{equation}
    q_n(t)=(\delta_{n+1}^+)^{-1}\mathcal{M}_{n+1,12}^{(1)}(t)=-(\delta_{n+1}^+)^{-1}\overline{\hat{\mathbf{M}}_{n+1,21}^{(1)}(t)}.
  \end{equation}
Substituting the asymptotic expansion \eqref{tMasy1} into Eq.\,\eqref{eq:mscra} and collecting the same power of $\lambda$, we find the matrix-valued coefficients of $\lambda^j (j=0,1,2)$ vanish from $\mathscr{A}_n(\lambda;t)$. Indeed,
\begin{equation}
\begin{aligned}
  \lambda^2:\quad& \begin{pmatrix}
  0&0\\\hat{\mathbf{M}}^{(0)}_{n+1,21}(t)&0
\end{pmatrix}=\mathbf{0},\\
  \lambda^1:\quad&\begin{pmatrix}
    0&0\\-\hat{\mathbf{M}}^{(1)}_{n+1,21}(t)&0
  \end{pmatrix}+\mathbf{Q}_n(t)\begin{pmatrix}
    \hat{\mathbf{M}}^{(0)}_{n+1,11}(t)&0\\ \hat{\mathbf{M}}^{(0)}_{n+1,21}(t)&0
  \end{pmatrix} \\
  &=\begin{pmatrix}
    0&0\\-\hat{\mathbf{M}}^{(1)}_{n+1,21}(t)-\delta_{n+1}^+\bar q_n(t)&0
  \end{pmatrix}=\mathbf{0},\\
  \lambda^0: \quad& \hat{\mathbf{M}}^{(0)}_{n}(t)- \hat{\mathbf{M}}^{(0)}_{n+1}(t)-\begin{pmatrix}
   0&0\\ \hat{\mathbf{M}}^{(2)}_{n+1,21}(t)&0
 \end{pmatrix}+\mathbf{Q}_n(t)\begin{pmatrix}
    \hat{\mathbf{M}}^{(1)}_{n+1,11}(t)&0\\ \hat{\mathbf{M}}^{(1)}_{n+1,21}(t)&0
 \end{pmatrix}\\
 &=\begin{pmatrix}
  \delta_n^+-\delta_{n+1}^--q_n(t)\delta_{n+1}^+\bar q_n(t) &0\\0&0
 \end{pmatrix}=\mathbf{0}.
\end{aligned}
\end{equation}
Therefore, $\mathscr{A}_n(t;\lambda)$   satisfies the homogeneous RH problem
\begin{subequations}
\begin{alignat}{2}
  &\mathscr{A}_{n+}(\lambda;t)=\mathscr{A}_{n-}(\lambda;t)\tilde{\mathbf{J}}_{n}(\lambda;t),\qquad&& \lambda\in\Sigma,\\
   & \mathscr{A}_n(\lambda;t)=O(\lambda^{-1}), \quad&& \lambda\rightarrow\infty,\label{eq:scraasy1}\\
    &    \mathscr{A}_n(\lambda;t)=O(\lambda), \quad&& \lambda\rightarrow 0, \label{eq:scraasy2}
  \end{alignat}
  \end{subequations}
  where Eq.\,\eqref{eq:scraasy2} is obtained directly from Eqs.\,\eqref{sym:mscrab} and \eqref{eq:scraasy1}.
 By virtue of  Lemma \ref{vanish}, we conclude
  \begin{equation}\label{LM}
    \mathscr{A}_n(\lambda;t)\equiv \mathbf{0}.
    \end{equation}
    Considering the coefficient of $\lambda$ and $\lambda^{-1}$ in the asymptotic expansions of $ \mathscr{A}_n(\lambda;t)$ as $\lambda\rightarrow\infty$ and  $\lambda\rightarrow 0$,  respectively, we find
    \begin{equation}
      q_n(t)=\overline{\mathcal{M}_{n,21}^{(1)}(t)}=-\hat{\mathbf{M}}_{n,12}^{(1)}(t).
    \end{equation}
Similar calculations yield that $\mathscr B_n(\lambda;t)$  satisfies the homogeneous RH problem
 \begin{alignat}{2}
   &\mathscr{B}_{n+}(t;\lambda)=\mathscr{B}_{n-}(t;\lambda)\tilde{\mathbf{J}}_n(\lambda;t),\qquad&& \lambda\in\Sigma,\nonumber\\
   & \mathscr{B}_n(\lambda;t)=O(\lambda^{-1}), \quad&& \lambda\rightarrow\infty,\\
    &    \mathscr{B}_n(\lambda;t)=O(\lambda), \quad&& \lambda\rightarrow 0,\nonumber
  \end{alignat}
which also implies
\begin{equation}\label{NM}
  \mathscr{B}_n(\lambda;t)\equiv \mathbf{0}.
  \end{equation}
  The compatibility condition of Eqs.\,\eqref{LM} and \eqref{NM}  yields the function $q_n(t)$ defined by Eq.\,\eqref{eq:recon1} solves the  Ablowitz--Ladik equation \eqref{eq:dfnls}.
\end{proof}
It follows from the jump condition \eqref{tMjump} that for $\lambda\in\Sigma$,
\begin{equation}
   \tilde{\mathbf{M}}_{n+}(\lambda;t)-  \tilde{\mathbf{M}}_{n-}(\lambda;t)=  \tilde{\mathbf{M}}_{n-}(\lambda;t)[\tilde{\mathbf{J}}_n(\lambda;t)-\mathbf{I}]=\tilde{\mathbf{M}}_{n+}(\lambda;t)[\mathbf{I}-\tilde{\mathbf{J}}^{-1}_n(\lambda;t)].
\end{equation}
Using the Plemelj--Sokhotski formula, we express $\tilde{\mathbf{M}}_n(\lambda;t)$ as an integral:
 \begin{equation}\label{eq:bfMn}
 \begin{split}
         \tilde{\mathbf{M}}_n(\lambda;t)
         =&\mathbf{I}+\frac{1}{2\pi\ii }\int_{\Gamma} \frac{\tilde{\mathbf{M}}_{n-}(\mu;t)[\mathbf{J}_n(\mu;t)-\mathbf{I}]}{\mu-\lambda}\dd \mu\\
         &+\sum_{k=1}^N\frac{1}{2\pi\ii }\int_{\partial \Omega_k} \frac{\tilde{\mathbf{M}}_{n-}(\mu;t)[\mathbf{P}_n^{-1}(k,\mu;t)-\mathbf{I}]}{\mu-\lambda}\dd \mu\\
         &+\sum_{k=1}^N\frac{1}{2\pi\ii }\int_{\partial \Omega_{-k}} \frac{\tilde{\mathbf{M}}_{n-}(\mu;t)[\mathbf{P}_n(k,-\mu;t)-\mathbf{I}]}{\mu-\lambda}\dd \mu\\
         &+\sum_{k=1}^N\frac{1}{2\pi\ii }\int_{\partial \tilde{\Omega}_{k}} \frac{\tilde{\mathbf{M}}_{n+}(\mu;t)[\mathbf{I}-\mathbf{P}_n^{\dag}(k,\bar\mu^{-1};t)]}{\mu-\lambda}\dd \mu\\
         &+\sum_{k=1}^N\frac{1}{2\pi\ii }\int_{\partial \tilde{\Omega}_{-k}} \frac{\tilde{\mathbf{M}}_{n+}(\mu;t)[\mathbf{I}-(\mathbf{P}_n^{\dag}(k,-\bar\mu^{-1};t))^{-1}]}{\mu-\lambda}\dd \mu.
 \end{split}
 \end{equation}

 \section{Reflectionless potential }\label{less}
 Denote
 \begin{equation}
   \tilde{f}(\lambda,\mu;n,t)=\frac{f(\mu;t)\theta_n(\mu)}{\mu-\lambda},
 \end{equation}
it follows that $\tilde{f}(\lambda,\mu;n,t)$ is  analytic of $\mu $ at $\lambda_{\pm k}$ when $\lambda$ does not lie in $\Omega_{\pm k}$ for any $1\leqslant k\leqslant N$.
We now reconstruct the potential $ q_n(t)$ explicitly in the reflectionless case, i.e., $\gamma(\lambda)=0$. In this case, there is no jump across the contour  $\Gamma$. As a consequence, Eq.\,\eqref{eq:bfMn} becomes
\begin{equation}
\begin{split}
         \tilde{\mathbf{M}}_n(\lambda;t)
         =&\sum_{k=1}^N\frac{1}{2\pi\ii }\int_{\partial \Omega_k} \frac{\tilde{\mathbf{M}}_{n-}(\mu;t)[\mathbf{P}_n^{-1}(k,\mu;t)-\mathbf{I}]}{\mu-\lambda}\dd \mu\\
         &+\sum_{k=1}^N\frac{1}{2\pi\ii }\int_{\partial \Omega_{-k}} \frac{\tilde{\mathbf{M}}_{n-}(\mu;t)[\mathbf{P}_n(k,-\mu;t)-\mathbf{I}]}{\mu-\lambda}\dd \mu\\
         &+\sum_{k=1}^N\frac{1}{2\pi\ii }\int_{\partial \tilde{\Omega}_{k}} \frac{\tilde{\mathbf{M}}_{n+}(\mu;t)[\mathbf{I}-\mathbf{P}_n^{\dag}(k,\bar\mu^{-1};t)]}{\mu-\lambda}\dd \mu\\
         &+\sum_{k=1}^N\frac{1}{2\pi\ii }\int_{\partial \tilde{\Omega}_{-k}} \frac{\tilde{\mathbf{M}}_{n+}(\mu;t)[\mathbf{I}-(\mathbf{P}_n^{\dag}(k,-\bar\mu^{-1};t))^{-1}]}{\mu-\lambda}\dd \mu.
 \end{split}
 \end{equation}
By virtue of  Cauchy's Residue Theorem and the definition \eqref{eq:Pndef} of $\mathbf{P}_{n}(k,\lambda;t)$, the inverse problem reduces to be an algebraic system:
  \begin{subequations}\label{eq:Malge}
 \begin{align}
    &\tilde{\mathbf{M}}_{n,1}(\lambda;t)\nonumber\\
    =&\begin{pmatrix}
                             1 \\
                             0
                           \end{pmatrix}-\sum_{k=1}^N\left[\underset{\mu=\lambda_k}{\res}\frac{f(\mu;t)\theta_n(\mu)\tilde{\mathbf{M}}_{n,2}(\mu;t)}{(\mu-\lambda)(\mu-\lambda_k)^{m_k+1}}
                           +\underset{\mu=-\lambda_k}{\res}\frac{-f(-\mu;t)\theta_n(-\mu)\tilde{\mathbf{M}}_{n,2}(\mu;t)}{(\mu-\lambda)(-\mu-\lambda_k)^{m_k+1}}\right]\nonumber\\
    =&\begin{pmatrix}
                             1 \\
                             0
                           \end{pmatrix}-\sum_{k=1}^N\left[\underset{\mu=\lambda_k}{\res}\frac{f(\mu;t)\theta_n(\mu)\tilde{\mathbf{M}}_{n,2}(\mu;t)}{(\mu-\lambda)(\mu-\lambda_k)^{m_k+1}}+
                           \underset{\mu=\lambda_k}{\res}\frac{f(\mu;t)\theta_n(\mu)\tilde{\mathbf{M}}_{n,2}(-\mu;t)}{(-\mu-\lambda)(\mu-\lambda_k)^{m_k+1}}\right]\label{eq:algeM1}\\
    =&\begin{pmatrix}
                             1 \\
                             0
                           \end{pmatrix}+\sum_{k=1}^N\frac{1}{m_k!}\partial_\mu^{m_k}
    \left.\left(\frac{f(\mu;t) \theta_n(\mu)\tilde{\mathbf{M}}_{n,2}(\mu;t)}{\lambda-\mu}\right)\right|_{\mu=\lambda_k}\nonumber\\
    &+\sum_{k=1}^N\frac{1}{m_k!}\partial_\mu^{m_k}
    \left.\left(\frac{f(\mu;t)\theta_n(\mu)\tilde{\mathbf{M}}_{n,2}(-\mu;t)}{\lambda+\mu}\right)\right|_{\mu=\lambda_k},\quad \lambda\notin\{\Omega_{\pm k}\}_{k=1}^N,\nonumber\\
& \tilde{\mathbf{M}}_{n,2}(\mu;t)\nonumber\\
 =&\begin{pmatrix}
                             0 \\
                             1
                           \end{pmatrix}+\sum_{k=1}^N\left[\underset{\nu=\bar\lambda_k^{-1}}{\res}\frac{\overline{f(\bar\nu^{-1};t)\theta_n(\bar\nu^{-1})}\tilde{\mathbf{M}}_{n,1}(\nu;t)}{(\nu-\mu)(\nu^{-1}-\bar\lambda_k)^{m_k+1}}
                           +\underset{\nu=-\bar\lambda_k^{-1}}{\res}\frac{-\overline{f(-\bar\nu^{-1};t)\theta_n(-\bar\nu^{-1})}\tilde{\mathbf{M}}_{n,1}(\nu;t)}{(\nu-\mu)(-\nu^{-1}-\bar\lambda_k)^{m_k+1}}\right]\nonumber\\
 =&\begin{pmatrix}
                             0 \\
                             1
                           \end{pmatrix}+\sum_{k=1}^N\left[\underset{\nu=\bar\lambda_k^{-1}}{\res}\frac{\overline{f(\bar\nu^{-1};t)\theta_n(\bar\nu^{-1})}\tilde{\mathbf{M}}_{n,1}(\nu;t)}{(\nu-\mu)(\nu^{-1}-\bar\lambda_k)^{m_k+1}}
                           +\underset{\nu=\bar\lambda_k^{-1}}{\res}\frac{\overline{f(\bar\nu^{-1};t)\theta_n(\bar\nu^{-1})}\tilde{\mathbf{M}}_{n,1}(-\nu;t)}{(-\nu-\mu)(\nu^{-1}-\bar\lambda_k)^{m_k+1}}\right]\nonumber\\
 =&\begin{pmatrix}
                             0 \\
                             1
                           \end{pmatrix}-\sum_{k=1}^N\left[\underset{\nu=\bar\lambda_k}{\res}\frac{\overline{f(\bar\nu;t)\theta_n(\bar\nu)}\tilde{\mathbf{M}}_{n,1}(\nu^{-1};t)}{\nu^2(\nu^{-1}-\mu)(\nu-\bar\lambda_k)^{m_k+1}}
                           +\underset{\nu=\bar\lambda_k}{\res}\frac{\overline{f(\bar\nu;t)\theta_n(\bar\nu)}\tilde{\mathbf{M}}_{n,1}(-\nu^{-1};t)}{\nu^2(-\nu^{-1}-\mu)(\nu-\bar\lambda_k)^{m_k+1}}\right]\label{eq:algeM2}\\
   = &\begin{pmatrix}
                             0 \\
                             1
                           \end{pmatrix}-\sum_{l=1}^N\frac{1}{m_l!}\partial_\nu^{m_l}
    \left.\left(\frac{\overline{f(\bar\nu;t)\theta_n(\bar\nu)}\tilde{\mathbf{M}}_{n,1}(\nu^{-1};t)}{(1-\mu\nu)\nu}\right)\right|_{\nu=\bar \lambda_l}\nonumber\\
    &+\sum_{l=1}^N\frac{1}{m_l!}\partial_\nu^{m_l}
    \left.\left(\frac{\overline{f(\bar\nu;t)\theta_n(\bar\nu)}\tilde{\mathbf{M}}_{n,1}(-\nu^{-1};t)}{(1+\mu\nu)\nu}\right)\right|_{\nu=\bar \lambda_l},\quad \mu\notin\{\tilde\Omega_{\pm k}\}_{k=1}^N.\nonumber
 \end{align}
 \end{subequations}

Let
\begin{align}
& h_n(\lambda;t)=\overline{f(\bar\lambda;t) \theta_n(\bar\lambda)},\\
&F_n(\lambda;t)=h_n(\lambda;t)\tilde{\mathbf{M}}_{n,11}(\lambda^{-1};t),\quad\quad
\tilde{F}_n(\lambda)=h_n(\lambda;t)\tilde{\mathbf{M}}_{n,11}(-\lambda^{-1};t),\\
&G_n(\lambda;t)= F_n(\lambda;t)-h_n(\lambda;t)\nonumber\\
    &+\sum_{k,l=1}^N\left[\frac{\partial_{\mu}^{m_k}\partial_{\nu}^{m_l}}{m_k!m_l!}\left.\left(\frac{h_n(\lambda;t)\overline{h_n(\bar\mu;t)}F_n(\nu;t)}{(\lambda^{-1}-\mu)(1-\mu\nu)\nu}\right)\right|_{\mu=\lambda_k\atop \nu=\bar\lambda_l}-\frac{\partial_{\mu}^{m_k}\partial_{\nu}^{m_l}}{m_k!m_l!}\left.\left(\frac{h_n(\lambda;t)\overline{h_n(\bar\mu;t)}\tilde{F}_n(\nu;t)}{(\lambda^{-1}-\mu)(1+\mu\nu)\nu}\right)\right|_{\mu=\lambda_k\atop \nu=\bar\lambda_l}\right.\\
    &+\left.\frac{\partial_{\mu}^{m_k}\partial_{\nu}^{m_l}}{m_k!m_l!}\left.\left(\frac{h_n(\lambda;t)\overline{h_n(\bar\mu;t)}F_n(\nu;t)}{(\lambda^{-1}+\mu)(1+\mu\nu)\nu}\right)\right|_{\mu=\lambda_k\atop \nu=\bar\lambda_l}-\frac{\partial_{\mu}^{m_k}\partial_{\nu}^{m_l}}{m_k!m_l!}\left.\left(\frac{h_n(\lambda;t)\overline{h_n(\bar\mu;t)}\tilde{F}_n(\nu;t)}{(\lambda^{-1}+\mu)(1-\mu\nu)\nu}\right)\right|_{\mu=\lambda_k\atop \nu=\bar\lambda_l}\right],\nonumber\\
    &\tilde{G}_n(\lambda;t)=\tilde{F}_n(\lambda;t)-h_n(\lambda;t)\nonumber\\
  &+\sum_{k,l=1}^N\left[\frac{\partial_{\mu}^{m_k}\partial_{\nu}^{m_l}}{m_k!m_l!}\left.\left(\frac{h_n(\lambda;t)\overline{h_n(\bar\mu;t)}F_n(\nu;t)}{(-\lambda^{-1}-\mu)(1-\mu\nu)\nu}\right)\right|_{\mu=\lambda_k\atop \nu=\bar\lambda_l}-\frac{\partial_{\mu}^{m_k}\partial_{\nu}^{m_l}}{m_k!m_l!}\left.\left(\frac{h_n(\lambda;t)\overline{h_n(\bar\mu;t)}\tilde{F}_n(\nu;t)}{(-\lambda^{-1}-\mu)(1+\mu\nu)\nu}\right)\right|_{\mu=\lambda_k\atop \nu=\bar\lambda_l}\right.\\
    &+\left.\frac{\partial_{\mu}^{m_k}\partial_{\nu}^{m_l}}{m_k!m_l!}\left.\left(\frac{h_n(\lambda;t)\overline{h_n(\bar\mu;t)}F_n(\nu;t)}{(-\lambda^{-1}+\mu)(1+\mu\nu)\nu}\right)\right|_{\mu=\lambda_k\atop \nu=\bar\lambda_l}-\frac{\partial_{\mu}^{m_k}\partial_{\nu}^{m_l}}{m_k!m_l!}\left.\left(\frac{h_n(\lambda;t)\overline{h_n(\bar\mu;t)}\tilde{F}_n(\nu;t)}{(-\lambda^{-1}+\mu)(1-\mu\nu)\nu}\right)\right|_{\mu=\lambda_k\atop \nu=\bar\lambda_l}\right].\nonumber
\end{align}

Substituting Eq.\,\eqref{eq:algeM1} into  Eq.\,\eqref{eq:algeM2}, we conclude that
\begin{align}
  &G_n(\lambda;t)=0,\quad (\lambda;n,t)\in\mathbb{C}\times\mathbb{Z}\times\mathbb{R}^+\ \mbox{and}\ \lambda\neq \pm\lambda_1^{-1},\ldots,\pm\lambda_N^{-1},\\
  &\tilde{G}_n(\lambda;t)=0,\quad (\lambda;n,t;)\in\mathbb{C}\times\mathbb{Z}\times\mathbb{R}^+\ \mbox{and}\ \lambda\neq \pm\lambda_1^{-1},\ldots,\pm\lambda_N^{-1}.
\end{align}

\begin{theorem}\label{Th:refthe}
   In the reflectionless case, the solution of the  Ablowitz--Ladik  equation \eqref{eq:dfnls}   can be expressed  by
   \begin{equation}
     q_n(t)= -\sum_{k=1}^N\frac{1}{m_k!}\left(F_{n+1}^{(m_k)}(\bar\lambda_k;t)+\tilde{F}_{n+1}^{(m_k)}(\bar \lambda_k;t)\right),
   \end{equation}
   where $\{F_n^{(j_k)}(\bar\lambda_k;t), \tilde{F}_n^{(j_k)}(\bar \lambda_k;t)\}_{k=1,\ldots, N\atop  j_k=0,\ldots,m_k}$ is the solution of the following algebraic system
    \begin{equation}\label{eq:algesys}
     \begin{cases}
      \frac{ G_n^{(j_1)}(\bar\lambda_1;t)}{j_1!}=0,\quad &j_1=0,\ldots,m_1,\\
      \qquad \qquad\vdots\\
        \frac{G_n^{(j_N)}(\bar\lambda_N;t)}{j_N!}=0,\quad &j_N=0,\ldots,m_N,\\
         \frac{\tilde{ G}_n^{(j_1)}(\bar \lambda_1;t)}{j_1!}=0,\quad &j_1=0,\ldots,m_1,\\
      \qquad \qquad\vdots\\
        \frac{\tilde{G}_n^{(j_N)}(\bar \lambda_N;t)}{j_N!}=0,\quad &j_N=0,\ldots,m_N.
     \end{cases}
   \end{equation}
\end{theorem}
Specially, let $N=1$, $\lambda_1=2\mathrm{i}$, $f(\lambda)=\lambda^2$, when $m_1=0$, we obtain the one-soliton solution
\begin{equation}
  q_n(t)=-\frac{450 \e^{\ii \pi  n+\frac{25 \ii t}{4}}}{225\times 4^n+4^{3-n}},
\end{equation}
when $m_1=1$, we obtain the second  order poles soliton solution
\begin{align}
  q_n(t)
 =&3375 \e^{\ii \pi  n+\frac{25 \ii t}{4}} \left\{759375\times 16^n (15 t+4\ii n)-1024\ii (60 n+225\ii t+136)\right\}/\nonumber\\
 & \left\{151875\times 2^{2 n+3} \left(80 n (15 n+34)+16875 t^2+1712\right)\right.\\
 &\left.+2562890625\times 2^{6 n+1}+2^{21-2 n}\right\},\nonumber
\end{align}
when $m_1=2$, we obtain the third  order poles soliton solution
\begin{align}
  q_n(t)
  =&759375\times 4^n \e^{\ii \pi  n+\frac{25\ii t}{4}}\nonumber \\
  &\left\{1946195068359375\times 256^n \left(-2\ii (60 n-19) t+8 n (2 n+1)-225 t^2\right)\right.\nonumber\\
  &+268435456 \left(30\ii (60 n+253) t+24 n (10 n+73)-3375 t^2+2990\right)\nonumber \\
  &-2278125\times 2^{4 n+5}\left(33750 (360 n (10 n+39)+14201) t^2\right.\nonumber\\
  &+240 n (10 n (360 n (5 n+39)+40291)+487253)\nonumber\\
  &\left.\left.-1500\ii (180 n (272 n+1157)+233863) t+854296875 t^4+44051488\right)\right\}/\\
    &\left\{1946195068359375\times 256^n \left(3037500 \left(8 n^2+4 n-3\right) t^2\right.\right.\nonumber\\
    &\left.+64 (3 n (15 n (20 n (15 n+83)+3663)+59618)+83663)+170859375 t^4\right)\nonumber\\
     &+34171875\times 4^{2 n+7}\left(810000 (15 n (10 n+73)+2053) t^2\right.\nonumber\\
     &\left.+4 (600 n (12 n (10 n (15 n+151)+5569)+105677)+36411859)+854296875 t^4\right)\nonumber\\
     &\left.+1477891880035400390625\times 4^{6 n+2}+70368744177664\right\}.\nonumber
  \end{align}

\begin{figure}[ht]
\centering
\begin{subfigure}[t]{0.35\textwidth}
\centering
\includegraphics[width=1\textwidth]{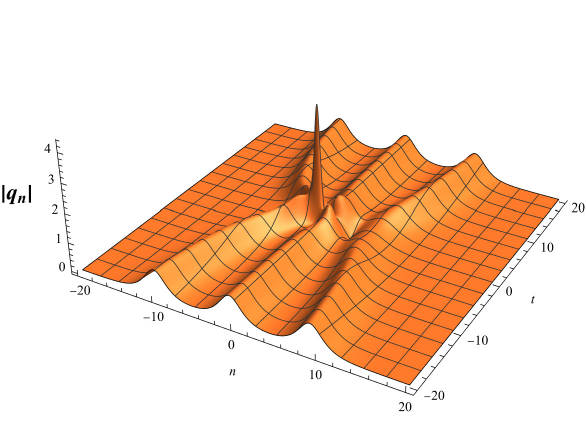}
\end{subfigure}
\quad
\begin{subfigure}[t]{0.35\textwidth}
\centering
\includegraphics[width=1\textwidth]{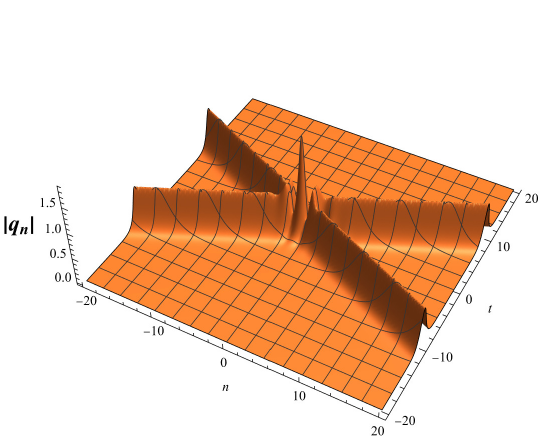}
\end{subfigure}
\caption{{\bf Left}: A three-order pole solution by choosing $N=1$, $\lambda_1=1.5\mathrm{i}$, $m_1=2$ and $f(\lambda)=\lambda^2$;  {\bf Right}: Two-soliton solution by choosing $N=2$, $m_1=m_2=0$, $\lambda_1=1-\mathrm{i}$, $\lambda_2=1+\mathrm{i}$ and $f(\lambda)=1$.}
\label{fig2}
\end{figure}

In the following, we display the diagrams of two explicit solutions by regrading the discrete variable $n$ as a continuous variable, see figure \ref{fig2}.

\section*{Acknowledgment}
This work was supported by National Natural Science Foundation of China (Grant Nos. 12171439, 12101190, 11931017).

\end{document}